\documentclass[a4paper,11pt]{article}

\usepackage{pos}
\usepackage{lipsum} 

\usepackage{xcolor}
\usepackage{float}
\usepackage{graphicx}
\usepackage{dcolumn}
\usepackage{bm}
\usepackage{hyperref}

\usepackage[normalem]{ulem}

\def\rbm#1{\xrbm#1\relax^\relax\valign}
\def\xrbm#1^#2\relax#3\valign{%
\mathbf{#1}\ifx\valign#2\valign\else^{\mathbf{#2}}\fi}

\usepackage{lineno}
\title{Reconstruction of cosmic-ray properties with uncertainty estimation using graph neural networks in GRAND}

\ShortTitle{Reconstruction of cosmic-ray properties with GNN in GRAND }

\author*[a,b]{Arsène Ferrière}
\author[a]{Aurélien Benoit-Lévy}
\onbehalf{for the GRAND Collaboration{\normalsize \\ \normalfont(a complete list of authors can be found at the end of the proceeding)}\\}

\affiliation[a]{Université Paris-Saclay, CEA, List, F-91120 Palaiseau, France}
\affiliation[b]{Sorbonne Université, Université Paris Diderot, Sorbonne Paris Cité, CNRS, Laboratoire de Physique 5 Nucléaire et de Hautes Energies (LPNHE), 6 4 place Jussieu, F-75252, Paris Cedex 5, France}

\emailAdd{arsene.ferriere@cea.fr}

\abstract{

The Giant Radio Array for Neutrino Detection (GRAND) aims to detect and study ultra-high-energy (UHE) neutrinos by observing the radio emissions produced in extensive air showers. The GRANDProto300 prototype primarily focuses on UHE cosmic rays to demonstrate the autonomous detection and reconstruction techniques that will later be applied to neutrino detection. In this work, we propose a method for reconstructing the arrival direction and energy with high precision using state-of-the-art machine learning techniques from noisy simulated voltage traces.

For each event, we represent the triggered antennas as a graph structure, which is used as input for a graph neural network (GNN). To significantly enhance precision and reduce the required training set size, we incorporate physical knowledge into both the GNN architecture and the input data. This approach achieves an angular resolution of 0.14° and a primary energy reconstruction resolution of about 15\%. Additionally, we employ uncertainty estimation methods to improve the reliability of our predictions. These methods allow us to quantify the confidence of the GNN predictions and provide confidence intervals for the direction and energy reconstruction.

Finally, we explore strategies to evaluate the consistency and robustness of the model when applied to real data. Our goal is to identify situations where predictions remain trustworthy despite domain shifts between simulation and reality.

\vspace{4mm}

}

\ConferenceLogo{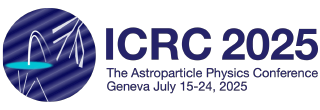}

\FullConference{39th International Cosmic Ray Conference (ICRC2025)\\  15--24 July, 2025\\ Geneva, Switzerland}

\begin{document}

\maketitle


\section{Introduction}
As the GRAND experiment \cite{GRAND_2019} starts measuring cosmic ray candidates through radio emission by the induced air shower at both the GRANDProto300 \cite{Proto300} site in the Gansu province in China and at the Pierre Auger Observatory site in Argentina~\cite{gaa}, reliable and precise reconstruction techniques for direction and energy are necessary.
The aim of this study is to use a Graph Neural Network (GNN) to reconstruct these two parameters based directly on voltage signals measured by the antennas. The GNN model takes as input a graph constructed from antenna positions and features extracted from voltage signals and outputs the desired quantity. It relies on both a data-driven approach where many simulations are used to train the model as well as a physics based method to serve as a first estimation of the targeted quantities.
The output of the GNN is then compared to the output from other methods to validate the reconstruction on cosmic ray candidates. The GNN approach demonstrates potential for good direction resolution when applied on measured voltage directly. This analysis provides an early demonstration of the capabilities of machine learning within the GRAND framework for cosmic ray parameters reconstruction.

\section{Methodology}
\subsection{Simulated Data}
To train the machine learning model, labeled data is required. For this reason, the training is conducted using simulated data. The simulations used are ZHaireS simulations \cite{Alvarez_Mu_iz_2012}, which are Monte Carlo simulations modeling the development of extensive air showers. These simulations provide realistic radio signals at the antenna level and simulate down-going proton- and iron-induced air showers with energies ranging from 0.4~EeV to 4~EeV. The zenith angle $\theta$ is defined as the angle between the shower axis and the vertical, while the azimuth angle $\phi$ is measured from geographic north, with 0° corresponding to north and 90° to west.

The electric field signals are then converted into voltage signals by modeling the full antenna response. This includes a detailed simulation of effective lengths of the antennas and their radio-frequency (RF) chain as functions of frequency, propagation direction, and signal polarization \cite{GRANDLIB_2025}. To increase the realism of the simulated data, noise sampled from real measurements taken at the GRANDProto300 site is added~\cite{nutrig}. This corresponds to colored noise with a standard deviation of $\sigma = 1.6\,\mathrm{mV}$. Additionally, a 5~ns time jitter is introduced to emulate uncertainties in the GPS time calibration, and a 7.5\% amplitude smearing is applied to account for fluctuations in gain and calibration errors.

For each antenna, only the time and amplitude of the peak of the voltage time trace are retained. These are extracted from the Hilbert envelope of the norm of the transverse (X, Y) components of the voltage signal. This provides a compact representation of the signal and helps minimize the simulation-to-reality gap by reducing dependence on detailed waveform features.

\begin{figure}[ht]
    \centering
    \includegraphics[width=1\linewidth]{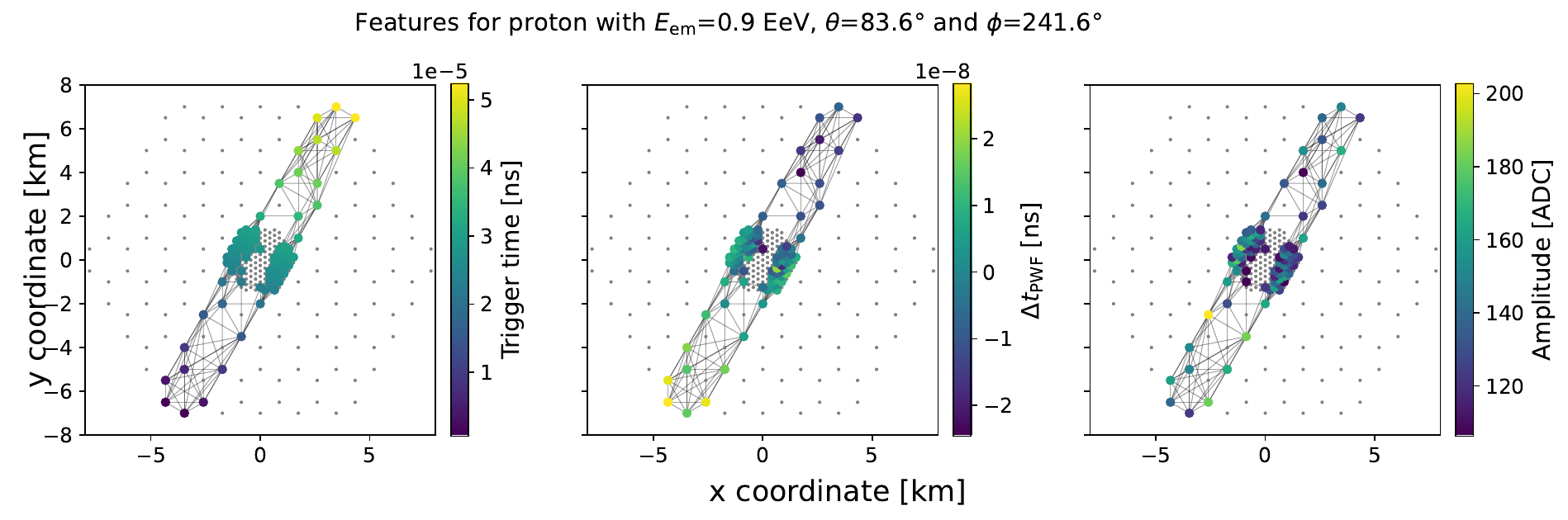}
    \caption{Example of a simulated cosmic-ray event represented as a graph. The spatial layout of the triggered antennas is shown in the left panel, with nodes corresponding to antenna positions in the array. From left to right, the color represent the following features: trigger time, $\Delta t_{\rm{PWF}}$ and peak amplitude. $E_{\rm{em}}$ is the energy in the electric field.}
    \label{fig:graph_and_features}
\end{figure}

A quality cut is applied to the data to model realistic triggering conditions. Antennas with low signal amplitudes would not trigger in practice, so we discard any antenna signal whose peak amplitude is below 5$\sigma$. This cut ensures that only realistic detections are retained and that the extracted peak values are not dominated by noise fluctuations. Additionally, we keep only events with more than five antennas remaining after the quality cut. In total, 7,400 simulated events are retained from an initial set of 33,000.
We also compute a planar wavefront fit (PWF) \cite{Ferriere_2025} and calculate the residual between the measured arrival times and the expected times assuming a planar wavefront. This quantity, $\Delta t_{\rm{PWF}}$, is used during training in some configurations described later.
The final step in data pre-processing is constructing the graph structure for the GNN. After testing multiple configurations, we opted to connect each antenna to its eight nearest neighbors. The resulting graph contains nodes representing antennas, each with 5 or 6 associated features: the X, Y, Z coordinates, peak time, amplitude, and optionally $\Delta t_{\rm{PWF}}$. An example of such an event graph is shown in Figure~\ref{fig:graph_and_features}.

\subsection{Graph Neural Network and training procedure}
We consider two closely related architectures: one purely data-driven, referred to as rGNN (raw GNN), and another that incorporates physical knowledge by using precomputed estimations of event-level quantities. In particular, the direction of propagation obtained from the PWF fit is used as additional input. These quantities are incorporated into the model at two different stages. This second model is referred to as pGNN, as it leverages the PWF information. Having both architectures (rGNN and pGNN) provides two independent reconstruction methods, which is valuable for cross-checking predictions and assessing confidence when applying the models to real data.

\begin{figure}[ht]
    \centering
    \includegraphics[width=1\linewidth]{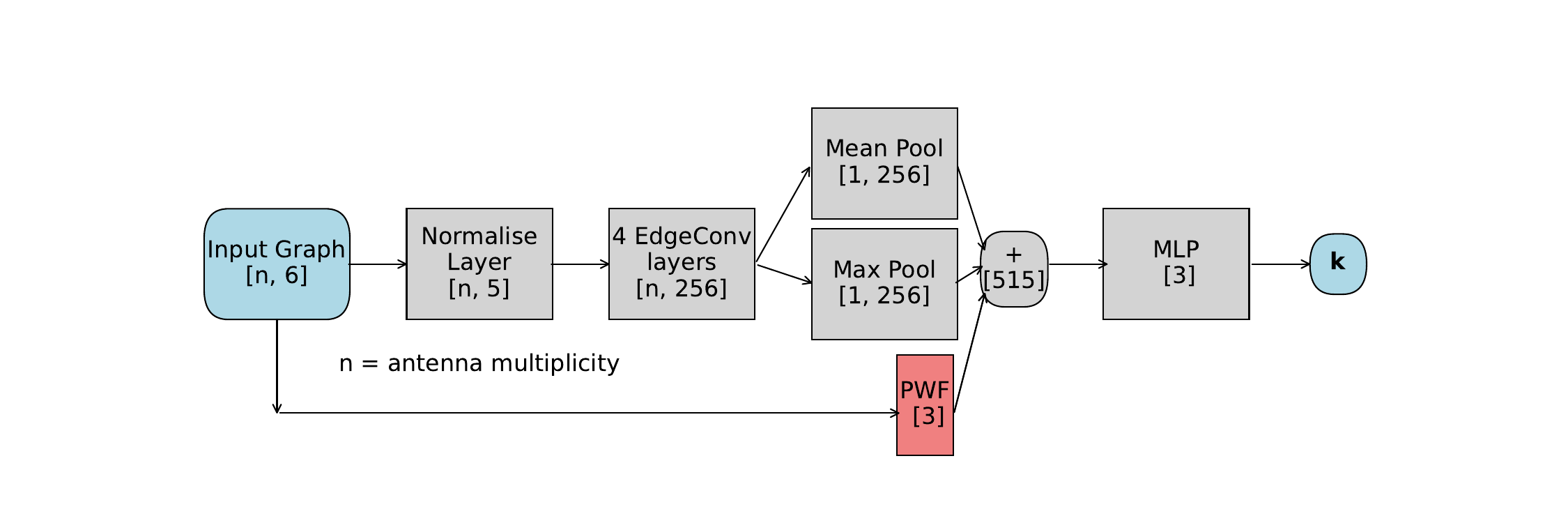}
    \caption{Architecture of the graph neural network. In rGNNs, the bottom branch is removed and the input vector is only of size 5.}
    \label{fig:GNN-architecture}
\end{figure}

The graph neural network takes as input a graph of $n$ nodes, each with a 5-dimensional feature vector (6 in the case of pGNN, which includes $\Delta t_{\rm{PWF}}$). After normalization, the features are passed through 2 to 4 EdgeConv layers, as defined in \cite{Edge_conv}, which capture the local spatial structure between antennas. The resulting node features are aggregated using mean and max pooling, producing a 512-dimensional embedding vector. For pGNN, this vector is concatenated with the PWF output, providing a physics-informed prior to guide the model. The final representation is passed through a Multi-Layer Perceptron (MLP), which outputs either a 3D direction vector $\rbm k$ or a scalar energy estimate. The full architecture is shown in Figure~\ref{fig:GNN-architecture}.

To train the network, the simulation dataset is split into two subsets: a training set containing 5,897 events and a validation set with 1,504 events. Training is performed over 110 epochs, during which the network is exposed to the entire training set in each epoch.

In practice, the models predict both the target quantity and the associated uncertainty, represented as the variance of the prediction, following a mixture density network–like approach \cite{bishop1994mixture}. The loss function that is minimized is:
\begin{equation}
\mathcal{L}_{\text{NLL}} = \frac{1}{2} \log \hat{\sigma}^2 + \frac{(y - \hat{y})^2}{2\hat{\sigma}^2},
\end{equation}
where $y$ is the ground truth value, $\hat{y}$ is the predicted value, and $\hat{\sigma}^2$ is the predicted variance. This loss encourages the model not only to make accurate predictions, but also to estimate realistic uncertainties: overconfident errors are penalized more strongly, while well-calibrated predictions are rewarded.
For vector-valued predictions, such as the 3D direction of the incoming particle, this loss is computed separately for each component and summed. As a result, the model predicts six values: three means and three variances.

To further improve robustness and capture model uncertainty, we train an ensemble of 30 independent models with different random initializations and data shuffling. At inference time, the final predicted value is taken as the mean of the 30 individual predictions. The total predicted uncertainty is obtained by combining the average predicted variance (aleatoric uncertainty) with the variance across the ensemble predictions (epistemic uncertainty).

\section{Performance on simulations}
To assess the reconstruction performance for direction, we study the zenith angle error and the angular resolution, defined as the 3D angle between the predicted and true directions. The performance of the models is compared with that of the planar wavefront fit.
For energy reconstruction, we evaluate the energy resolution, defined as $(\hat{E}_{\rm{em}} - E_{\rm{em}})/{E_{\rm{em}}}$, expressed as a percentage, where $E_{\rm{em}}$ is the electromagnetic energy of the shower.

\begin{figure}[ht]
    \centering
    \includegraphics[width=.6\linewidth]{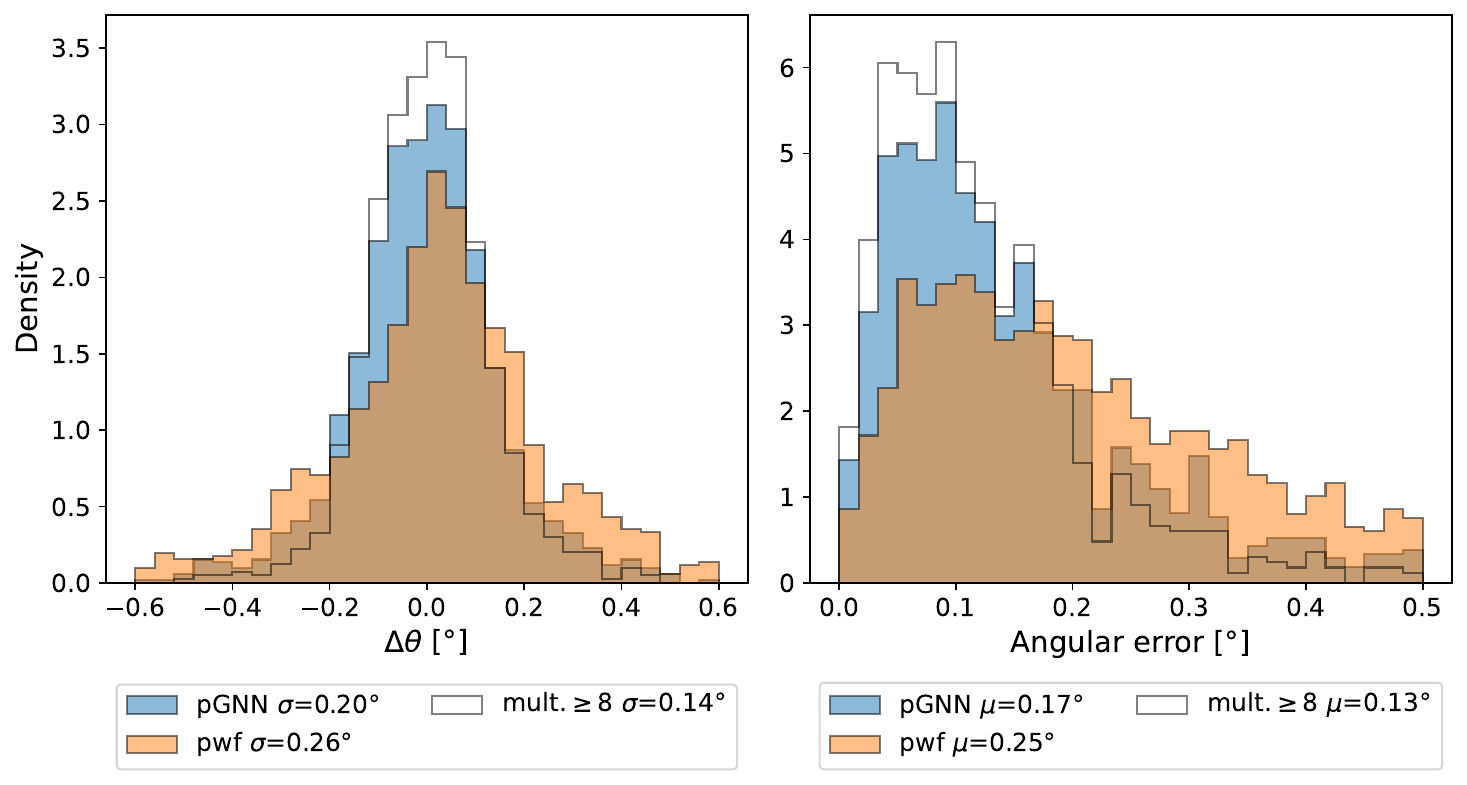}
    \includegraphics[width=.39\linewidth]{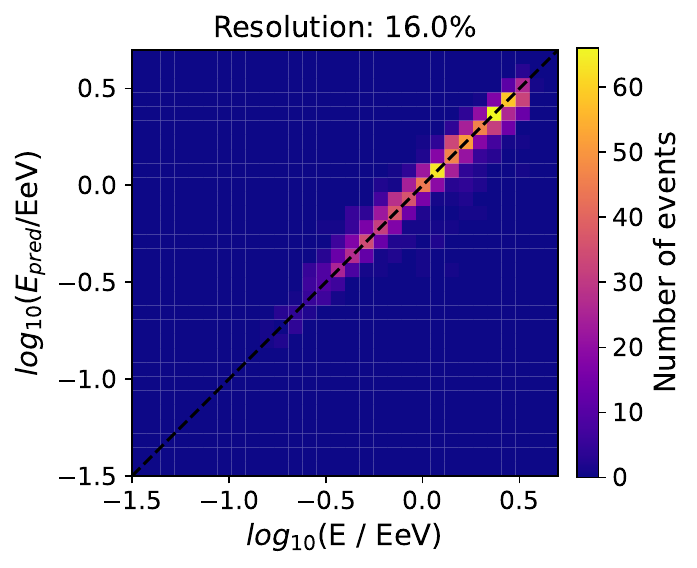}
    \caption{{\bf Left} and {\bf middle:} Distribution of the zenith angle residuals (left panel) and the angular errors (middle panel) for the PWF fitting method (orange), the pGNN (blue), and pGNN keeping events with an antenna multiplicity of at least 8 antennas (solid black line). $\sigma$ is the standard deviation of the residuals on $\theta$, $\mu$ is the mean angular error: the angular resolution.  For comparison, the rGNN achieves an angular resolution of 0.9°. {\bf Right:} Energy reconstruction resolution, showing the correlation between predicted and true electromagnetic energy.
}
    \label{fig:avp}
\end{figure}

As shown in Figure~\ref{fig:avp}, the pGNN achieves an angular resolution of 0.17°. This resolution depends strongly on the number of triggered antennas. When restricting to events with more than 8 antennas, it improves to 0.13°.

For energy reconstruction, the average resolution is 16\%, and remains approximately constant across the full energy range from 0.4~EeV to 4.0~EeV. However, the precision improves for less inclined events: for zenith angles below 83°, the resolution drops to 13.6\%.

\begin{figure}[ht]
    \centering
    \includegraphics[width=0.45\linewidth]{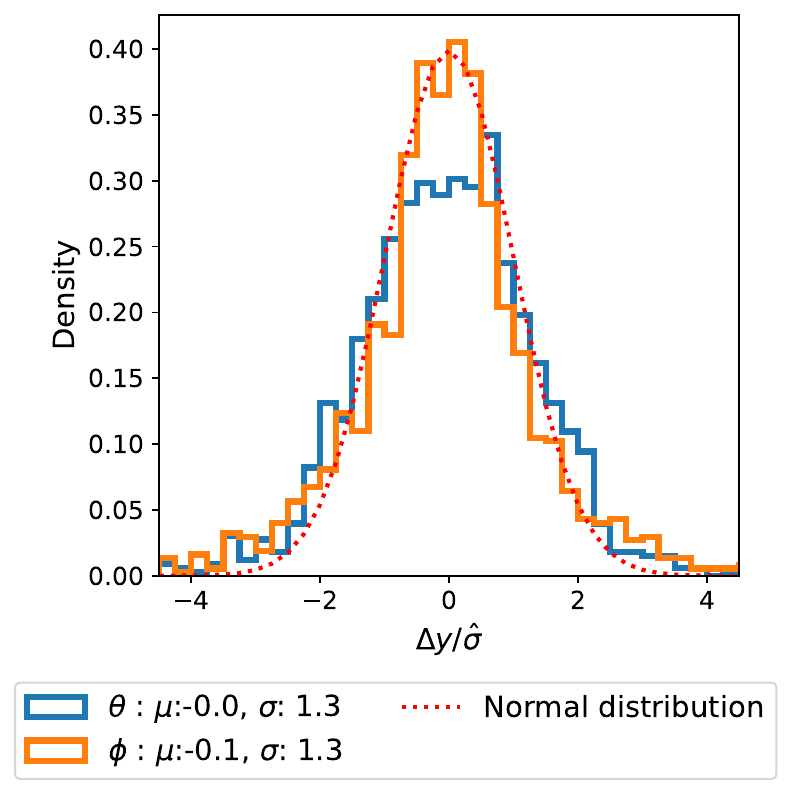}
    \includegraphics[width=0.45\linewidth]{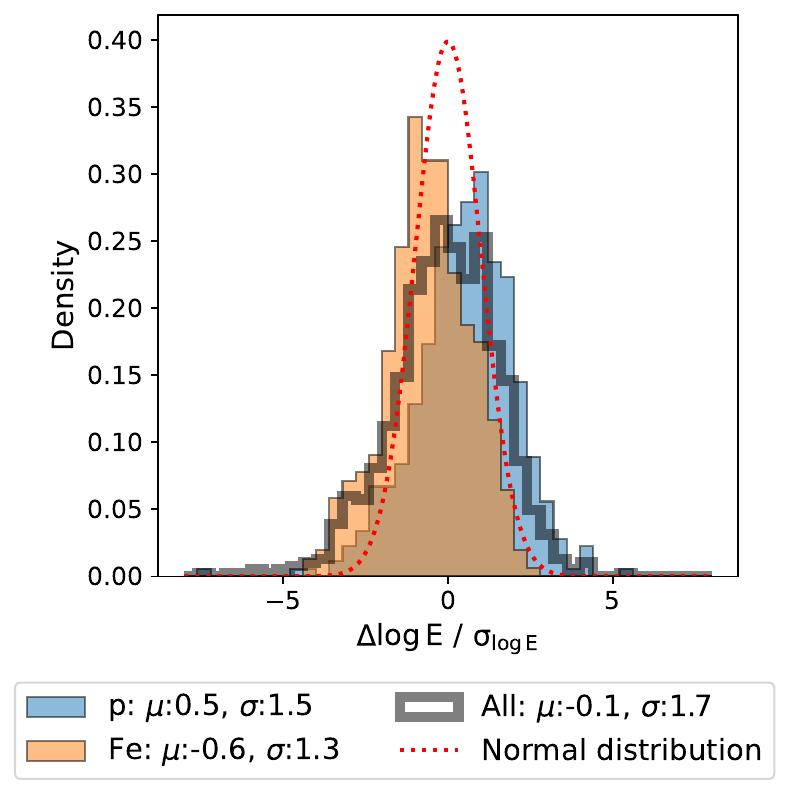}
    \caption{Distribution of normalized residuals. {\bf Left:} Zenith (blue) and azimuth (orange) reconstruction. {\bf Right:} Energy reconstruction, separated by primary type.}
\label{fig:uncertainty_calibration}
\end{figure}

To assess uncertainty calibration, we evaluate the distribution of normalized residuals, defined as $\Delta y / \hat{\sigma} = (y - \hat{y}) / \hat{\sigma}_{\rm{total}}$. If the predicted uncertainties are well calibrated and the errors follow a Gaussian distribution, the normalized residuals should follow a standard normal distribution $\mathcal{N}(0, 1)$. These distributions for both direction and energy are shown in Figure~\ref{fig:uncertainty_calibration}. In all cases, the uncertainties are underestimated by a factor of about 1.5. However, the distributions remain Gaussian, indicating that the uncertainty estimates can still serve as a reasonable first-order approximation.

\section{Reconstruction of cosmic-ray candidates}

We have shown that our machine learning approach performs well on simulated data. Here, we apply the trained models to the 40 cosmic ray candidates from GRANDProto300 identified by the collaboration, as described in \cite{CRsearch}. This introduces potential mismatches between data and simulation. Differences in antenna calibration, trigger algorithms, or the nature of the events (e.g., anthropogenic noise or different primaries) may cause real inputs to deviate from the training distribution, potentially affecting the reliability of the GNN predictions.

\subsection{Assessing confidence}
Evaluating the reliability of predictions from graph neural networks is not straightforward, as there is no direct way to assess fit quality—unlike in traditional likelihood-based reconstruction. To address this, we adopt an indirect validation approach: we compare the distributions of differences between several reconstruction methods, using both simulated and measured data. If these distributions are similar, it suggests that the models behave consistently on real data, increasing our confidence in their predictions. This strategy is applied to direction reconstruction, where we can rely on multiple independent methods (PWF, pGNN, and rGNN) for comparison.

We also verify that the distributions of voltage amplitudes, signal arrival times, and the PWF residuals $\Delta t_{\rm{PWF}}$ are similar between simulations and candidate events. This indicates that the data do not lie significantly out of distribution compared to the simulations, and supports the safe application of the machine learning models to real data.

\begin{figure}[ht]
    \centering
    \includegraphics[width=0.6\linewidth]{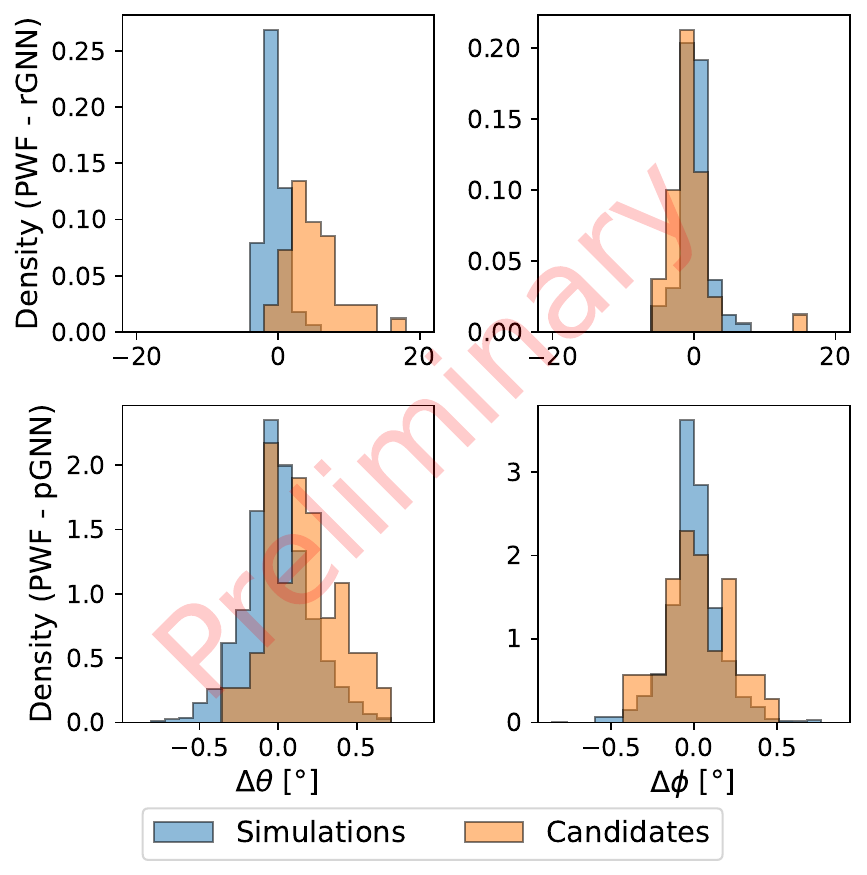}
    \caption{Comparison of reconstruction differences between methods for simulations and measured candidate events. The \textbf{left panels} show the distribution of angular differences in $\theta$ between GNNs and PWF, and the \textbf{right panels} show the same for $\phi$. The \textbf{top row} correspond to comparison between rGNN and PWF, the \textbf{bottom row} between pGNN and PWF. Simulation events were selected with $\theta > 70^\circ$ and multiplicity $< 10$ to match the data distribution.}
    \label{fig:difference_reconstruction}
\end{figure}
However, when performing inference using pGNN, rGNN and examining the distributions of reconstruction differences with PWF, as shown in Figure~\ref{fig:difference_reconstruction}, we observe a discrepancy in the zenith angle for rGNN while the azimuth angle remains consistent across all methods. For pGNN, the distributions differ less significantly. These discrepancies can be explained by differences in trigger patterns. For a cosmic ray with fixed energy, zenith, and azimuth, fewer antennas are triggered on-site than in simulation, due to the smaller array size and inactive antennas in the real detector. The GNNs appear to have learned a correlation between lower multiplicity and more vertical showers. This hypothesis can be verified in simulation by reducing the array layout to match the one deployed on site and randomly removing 50\% of the remaining antennas. We then observe the same bias toward lower zenith angles.
\subsection{Reconstruction}

\begin{figure}[ht]
\begin{minipage}[t]{0.5\linewidth}
\begin{center}
\includegraphics[width=0.99\linewidth]{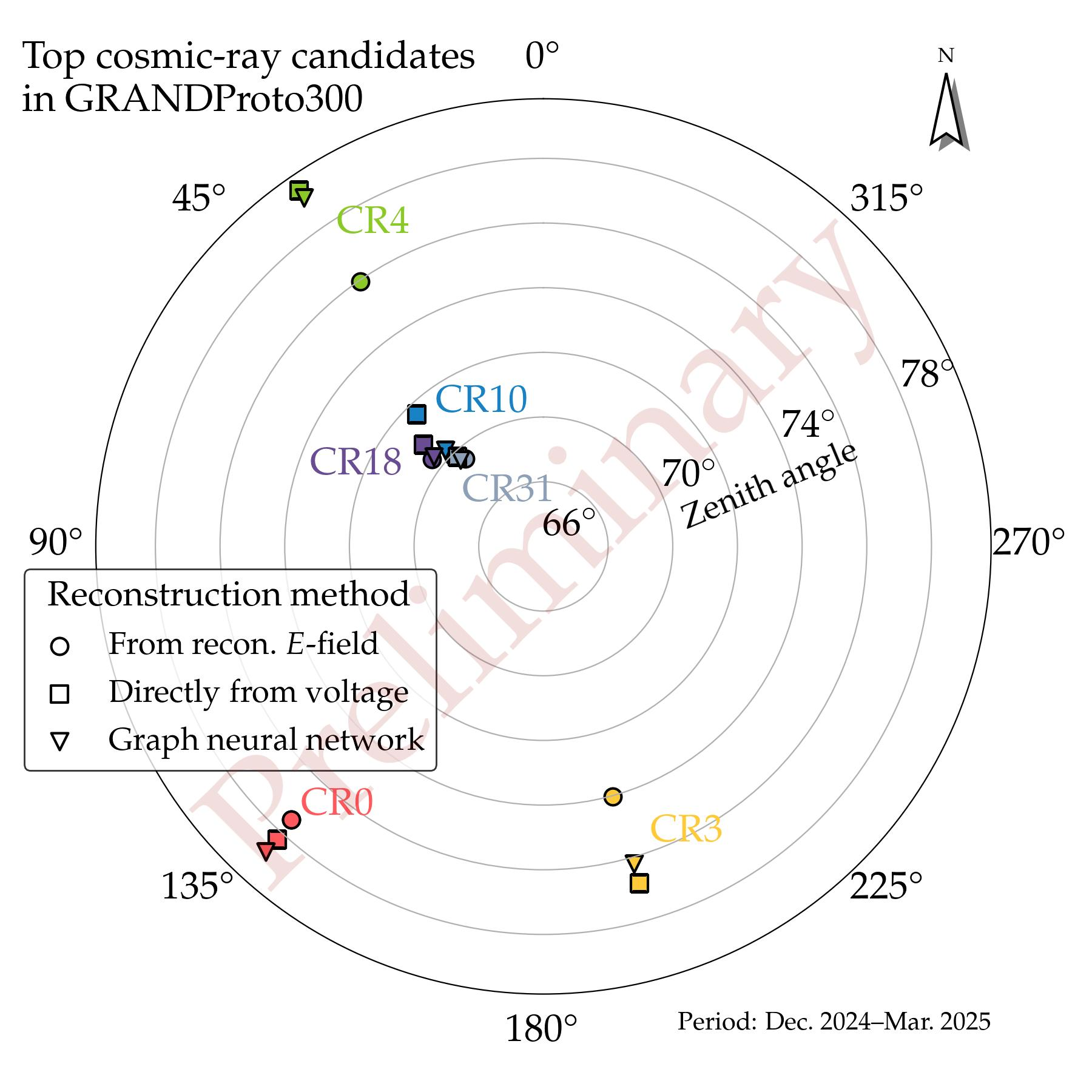}
\end{center}
\end{minipage}
\begin{minipage}[t]{0.5\linewidth}
\begin{center}
\includegraphics[width=0.99\linewidth]{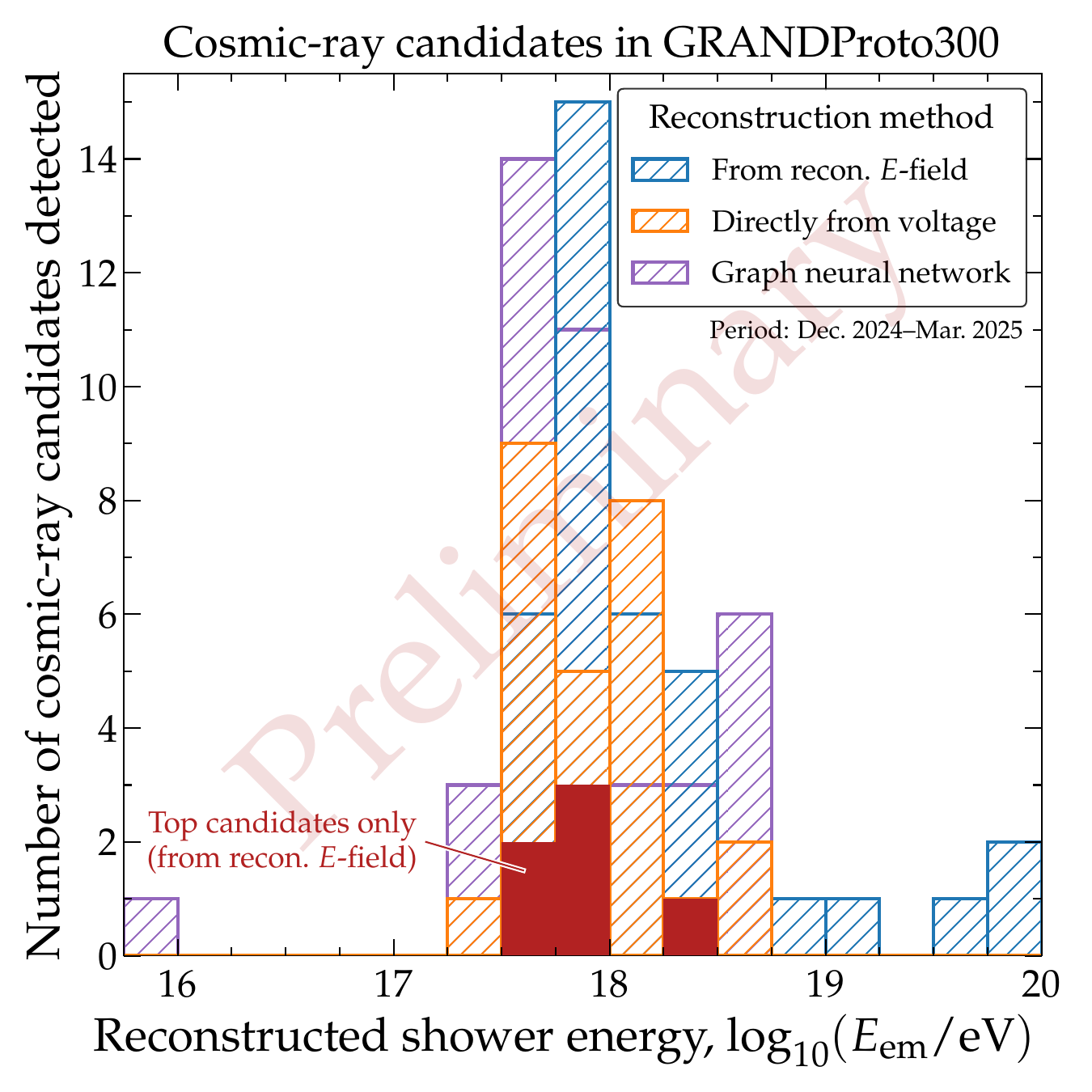}
\end{center}
\end{minipage}
\caption{\label{fig:energy_predictions}{\bf Left:} Polar plot showing the reconstructed arrival directions of top cosmic-ray candidate events using the pGNN model. A comparison with other reconstruction methods is presented in related contributions. {\bf Right:} Distribution of reconstructed electromagnetic-equivalent shower energy using different reconstruction methods.}
\end{figure}

Despite the observed inconsistencies, we can still estimate the arrival directions of candidate events using the pGNN approach. As shown in Figure~\ref{fig:energy_predictions} (left panel), the reconstructed directions for the top cosmic-ray candidates are displayed in polar coordinates. The results show strong consistency in azimuth and reasonably good agreement in zenith angle when compared to the voltage-based reconstruction method described in \cite{ADF}. This comparison supports the reliability of pGNN predictions, even in the low-multiplicity regime.

As for the energy reconstruction, the predicted energy distribution obtained with our method shows strong agreement with those from other reconstruction approaches from \cite{LDF, ADF}, as illustrated in Figure~\ref{fig:energy_predictions} (right panel). The consistency in the overall shape and peak position of the distributions suggests that the models yield comparable energy scales and are sensitive to similar underlying shower characteristics.

\section{Conclusion}

We have presented a machine learning-based approach for reconstructing the direction and energy of cosmic rays using graph neural networks, trained on realistic Monte-Carlo simulations. The method incorporates physical information, uncertainty estimation, and ensemble techniques to achieve high reconstruction accuracy and reliability.

On simulations, the pGNN model achieves an angular resolution better than 0.2°, and an energy resolution below 16\%, with close to calibrated uncertainty estimates. Applying the model to measured data from the GRANDProto300 setup shows promising agreement with traditional reconstruction methods in both direction and energy. While some discrepancies remain, the overall consistency between simulation and data supports the validity of the approach.

This work demonstrates the potential of graph-based deep learning to enhance reconstruction capabilities in the GRAND project, paving the way for robust and scalable inference methods in future large-scale radio arrays.

\bibliographystyle{ICRC}
\bibliography{references}

\clearpage

\section*{Full Author List: GRAND Collaboration}

\scriptsize
\noindent
J.~Álvarez-Muñiz$^{1}$, R.~Alves Batista$^{2, 3}$, A.~Benoit-Lévy$^{4}$, T.~Bister$^{5, 6}$, M.~Bohacova$^{7}$, M.~Bustamante$^{8}$, W.~Carvalho$^{9}$, Y.~Chen$^{10, 11}$, L.~Cheng$^{12}$, S.~Chiche$^{13}$
% ...
\\
\\
$^{1}$Departamento de Física de Particulas \& Instituto Galego de Física de Altas Enerxías, Universidad de Santiago de Compostela, 15782 Santiago de Compostela, Spain \\
$^{2}$Institut d'Astrophysique de Paris, CNRS  UMR 7095, Sorbonne Université, 98 bis bd Arago 75014, Paris, France \\
$^{3}$Sorbonne Université, Université Paris Diderot, Sorbonne Paris Cité, CNRS, Laboratoire de Physique  Nucléaire et de Hautes Energies (LPNHE), 4 Place Jussieu, F-75252, Paris Cedex 5, France \\
$^{4}$Université Paris-Saclay, CEA, List,  F-91120 Palaiseau, France \\
$^{5}$Institute for Mathematics, Astrophysics and Particle Physics, Radboud Universiteit, Nijmegen, the Netherlands \\
$^{6}$Nikhef, National Institute for Subatomic Physics, Amsterdam, the Netherlands \\
$^{7}$Institute of Physics of the Czech Academy of Sciences, Na Slovance 1999/2, 182 00 Prague 8, Czechia \\
$^{8}$Niels Bohr International Academy, Niels Bohr Institute, University of Copenhagen, 2100 Copenhagen, Denmark \\
$^{9}$Faculty of Physics, University of Warsaw, Pasteura 5, 02-093 Warsaw, Poland \\
$^{10}$Key Laboratory of Dark Matter and Space Astronomy, Purple Mountain Observatory, Chinese Academy of Sciences, 210023 Nanjing, Jiangsu, China \\
$^{11}$School of Astronomy and Space Science, University of Science and Technology of China, 230026 Hefei Anhui, China \\
$^{12}$National Astronomical Observatories, Chinese Academy of Sciences, Beijing 100101, China \\
$^{13}$Inter-University Institute For High Energies (IIHE), Université libre de Bruxelles (ULB), Boulevard du Triomphe 2, 1050 Brussels, Belgium \\
% ...


\subsection*{Acknowledgments}

\noindent
The GRAND Collaboration is grateful to the local government of Dunhuag during site survey and deployment approval, to Tang Yu for his help on-site at the GRANDProto300 site, and to the Pierre Auger Collaboration, in particular, to the staff in Malarg\"ue, for the warm welcome and continuing support.
The GRAND Collaboration acknowledges the support from the following funding agencies and grants.
\textbf{Brazil}: Conselho Nacional de Desenvolvimento Cienti\'ifico e Tecnol\'ogico (CNPq); Funda\c{c}ão de Amparo \`a Pesquisa do Estado de Rio de Janeiro (FAPERJ); Coordena\c{c}ão Aperfei\c{c}oamento de Pessoal de N\'ivel Superior (CAPES).
\textbf{China}: National Natural Science Foundation (grant no.~12273114); NAOC, National SKA Program of China (grant no.~2020SKA0110200); Project for Young Scientists in Basic Research of Chinese Academy of Sciences (no.~YSBR-061); Program for Innovative Talents and Entrepreneurs in Jiangsu, and High-end Foreign Expert Introduction Program in China (no.~G2023061006L); China Scholarship Council (no.~202306010363); and special funding from Purple Mountain Observatory.
\textbf{Denmark}: Villum Fonden (project no.~29388).
\textbf{France}: ``Emergences'' Programme of Sorbonne Universit\'e; France-China Particle Physics Laboratory; Programme National des Hautes Energies of INSU; for IAP---Agence Nationale de la Recherche (``APACHE'' ANR-16-CE31-0001, ``NUTRIG'' ANR-21-CE31-0025, ANR-23-CPJ1-0103-01), CNRS Programme IEA Argentine (``ASTRONU'', 303475), CNRS Programme Blanc MITI (``GRAND'' 2023.1 268448), CNRS Programme AMORCE (``GRAND'' 258540); Fulbright-France Programme; IAP+LPNHE---Programme National des Hautes Energies of CNRS/INSU with INP and IN2P3, co-funded by CEA and CNES; IAP+LPNHE+KIT---NuTRIG project, Agence Nationale de la Recherche (ANR-21-CE31-0025); IAP+VUB: PHC TOURNESOL programme 48705Z. 
\textbf{Germany}: NuTRIG project, Deutsche Forschungsgemeinschaft (DFG, Projektnummer 490843803); Helmholtz—OCPC Postdoc-Program.
\textbf{Poland}: Polish National Agency for Academic Exchange within Polish Returns Program no.~PPN/PPO/2020/1/00024/U/00001,174; National Science Centre Poland for NCN OPUS grant no.~2022/45/B/ST2/0288.
\textbf{USA}: U.S. National Science Foundation under Grant No.~2418730.
Computer simulations were performed using computing resources at the CCIN2P3 Computing Centre (Lyon/Villeurbanne, France), partnership between CNRS/IN2P3 and CEA/DSM/Irfu, and computing resources supported by the Chinese Academy of Sciences.

\end{document}